\documentclass[a4paper,fleqn]{cas-sc}

\usepackage[authoryear]{natbib}
\usepackage{caption}

\def\tsc#1{\csdef{#1}{\textsc{\lowercase{#1}}\xspace}}
\tsc{WGM}
\tsc{QE}
\tsc{EP}
\tsc{PMS}
\tsc{BEC}
\tsc{DE}

\begin{document}
\let\WriteBookmarks\relax
\def\floatpagepagefraction{1}
\def\textpagefraction{.001}
\shorttitle{ATIM: Predicting Operator Action Time}
\shortauthors{Xiao et~al.}

\title{ATIM: An ACT-R-Based Task Interface Model for Predicting Operator Action Time in Digital Nuclear Control Rooms}

\tnotetext[1]{The research was supported by a grant from  the National Natural Science Foundation of China (Grant No. T2192933), the LingChuang Research Project of China National Nuclear Corporation, the Foundation of National Key Laboratory of Human Factors Engineering (Grant No. HFNKL2024W07), and Tsinghua University Initiative Scientific Research Program.}

\author[1,2]{Xingyu Xiao}[style=chinese]

\credit{Conceptualization, Methodology, Software, Formal analysis, Data Curation, Visualization, Validation, Writing- Original draft preparation.}

\author[3]{Jonghyun Kim}[style=chinese]
\credit{ Validation, Writing - Review and Editing.}

\author[1,2]{Jiejuan Tong}[style=chinese]
\credit{Conceptualization, Formal analysis, Supervision, Writing - Review and Editing.}

\author[1]{Jingang Liang}[orcid=0000-0003-2632-8613]
\cortext[cor1]{Corresponding author}
\cormark[1]
\ead{jingang@tsinghua.edu.cn; +86-10-62784836}
\credit{Supervision, Writing - review and editing.}

\author[1]{Haitao Wang}[style=chinese]
\credit{Supervision, Writing- Reviewing and Editing.}

\affiliation[1]{organization={Institute of Nuclear and New Energy Technology, Tsinghua University},
            city={Beijing},
            postcode={100084}, 
            country={China}}
\affiliation[2]{organization={National Key Laboratory of Human Factors Engineering},
            city={Beijing},
            postcode={100094}, 
            country={China}}

\affiliation[3]{organization={Korea Advanced Institute of Science and Technology},
            city={Daejeon},
            postcode={34141}, 
            country={Republic of Korea}}

\begin{abstract}
Human performance in digital control rooms is strongly influenced by interface characteristics, which shape visual search, cognitive processing, and motor execution. Accurate prediction of operator action time is therefore essential for ergonomic evaluation, interface design, and performance optimization in safety-critical systems. However, existing approaches typically rely on extensive experimental data or black-box models, limiting their interpretability and practical applicability. This study proposes ATIM (ACT-R-based Task Interface Model), a theory-guided and data-calibrated modeling framework that predicts operator action time directly from interface features. The model decomposes total action time into visual, cognitive, motor, and interaction components, integrating principles from visual search theory, Fitts’ law, and cognitive architecture modeling. Interface characteristics, including target salience and semantic interference, are used as direct inputs, enabling prediction without additional user experiments. A dataset from a digital nuclear control room environment was used for calibration and validation. Task types were identified through data-driven clustering, and separate parameter sets were estimated for novice and experienced users. The proposed framework achieved a mean absolute error of 3.17 s and demonstrated strong correlation with observed performance (r = 0.664) on a held-out validation set. The results show that ATIM generalizes well to unseen data while maintaining interpretability, offering a novel tool for ergonomic assessment and interface design in complex human–machine systems.

\end{abstract}


\maketitle

\section{Introduction}
Human performance plays a critical role in ensuring the safety and efficiency of nuclear power plant operations, particularly in digital main control rooms where human–machine interaction is highly dependent on interface design \cite{xiao2024emergency}. With the increasing digitalization of control systems, operators are required to process large volumes of information through complex graphical interfaces, perform visual search tasks, interpret semantic cues, and execute actions under time constraints \cite{xiao2025autograph, xiao2025krail}. In such environments, interface characteristics directly shape cognitive processes and behavioral outcomes, making them a key determinant of operator performance \cite{xiao2025insight}.

Among various performance indicators, operator action time is of particular importance because it is directly related to the time-reliability component of human reliability analysis \cite{zhu2019more}. In many HRA methods, the probability of successful human action depends not only on whether the operator selects the correct action, but also on whether the required action can be completed within the available time window \cite{leveson2004systems}. In this sense, operator action time provides a quantitative estimate of the required time demand, whereas the accident progression or operational procedure defines the available time. If the predicted action time approaches or exceeds the available time, the likelihood of delayed response, incomplete execution, or time-related human failure may increase substantially \cite{muller2019process}.

Therefore, accurate prediction of operator action time is not merely an ergonomic or efficiency issue, but a necessary basis for time-dependent HRA. In digital control rooms, interface characteristics such as target salience, semantic interference, and interaction distance can directly affect the time required for visual search, cognitive discrimination, and motor execution. These interface-induced delays may reduce the effective time margin available to operators during transient or emergency scenarios. Consequently, a model capable of predicting action time from measurable interface features can support both interface optimization and quantitative assessment of time-related human reliability.

Existing approaches for modeling operator performance typically fall into two categories. Data-driven methods, including machine learning models, can achieve high predictive accuracy but often rely on large-scale experimental datasets and lack interpretability, making them difficult to generalize across scenarios \cite{strielkowski2023prospects}. On the other hand, traditional human performance models, such as GOMS or cognitive architectures, provide mechanistic insights but are often limited in their ability to incorporate detailed interface characteristics as direct inputs \cite{olson1995growth}.

More importantly, most existing models do not explicitly capture the causal pathway from interface features to action time. In digital control rooms, performance is not solely determined by task complexity or operator expertise, but emerges from the interaction between interface design, cognitive processing, and motor execution. The lack of an interface-driven and theory-grounded modeling framework limits the applicability of current approaches in supporting interface design and real-time performance assessment \cite{horvath2023investigating}.

To address these limitations, this study proposes ATIM (ACT-R-based Task Interface Model), a theory-guided and data-calibrated modeling framework for predicting operator action time directly from interface features. The proposed framework is built upon the ACT-R cognitive architecture and decomposes total action time into multiple components, including visual search, cognitive processing, motor execution, and interaction dynamics. By integrating principles from visual search theory, Fitts’ law, and cognitive modeling, ATIM establishes a structured mapping from interface characteristics to time-based performance outcomes.

Unlike existing approaches, ATIM treats interface features, such as target salience, layout complexity, and semantic interference, as direct model inputs, enabling prediction without additional user experiments. This interface-driven formulation allows the model to capture the underlying mechanisms through which interface design influences operator behavior, thereby enhancing both interpretability and practical applicability.

The proposed framework is calibrated and validated using data collected from a digital nuclear control room environment. Task types are identified through data-driven clustering, and separate parameter sets are estimated for novice and experienced users to account for differences in cognitive processing and interaction strategies. The model performance is evaluated in terms of prediction accuracy and correlation with observed action time, demonstrating its capability to capture both absolute performance levels and relative variability across tasks.

This study makes three main contributions. First, it introduces an interface-driven and theory-grounded framework for predicting operator action time, bridging the gap between interface design and human performance modeling. Second, it extends ACT-R-based modeling by incorporating explicit representations of interface features, enabling more realistic and interpretable predictions. Third, it provides a practical tool for ergonomic assessment and interface optimization in digital control rooms, with potential applications in broader human–machine systems.

Unlike prior approaches, ATIM does not treat interface variables as generic predictors, but explicitly embeds them into a cognitively structured time decomposition, thereby establishing a direct mapping between interface design and human performance mechanisms.

\section{Related Work}\label{Related Work}

\subsection{Interface design and human performance} \label{Interface design and human performance}

Interface design plays a critical role in shaping human performance in complex human–machine systems. In modern digital control environments \cite{cooper2014face}, operators are required to process large volumes of information, locate relevant interface elements, and execute actions through structured graphical interfaces. Compared with traditional analog panels, digital interfaces introduce higher information density, multi-layer navigation, and software-mediated feedback, which increase both cognitive and interaction complexity \cite{crestani2004graphical}. As a result, operators must continuously allocate attention across multiple visual elements, interpret system states, and perform sequential operations under time pressure \cite{wickens2021attention}. These characteristics make interface design a central concern in ergonomics and human factors engineering, as even small inefficiencies in interaction may influence task performance in measurable ways \cite{karwowski2005ergonomics}.

Prior research has shown that specific interface characteristics can systematically affect different components of human performance. Visual layout and target salience are key determinants of search efficiency \cite{still2020examining}, with low-salience targets and cluttered displays leading to prolonged visual scanning. Information organization and semantic structure influence cognitive processing, as ambiguous or highly similar labels can introduce interference and increase decision latency \cite{rajeswar2022multi}. Furthermore, the spatial distribution of interface elements affects motor execution, since greater distances between interactive components generally result in longer movement times, consistent with established motor control principles. These findings highlight that visual, cognitive, and motor demands are all shaped by interface design and jointly contribute to overall task performance \cite{zago2009visuo}.

Importantly, the effects of these factors are not independent, but often interact in complex and non-linear ways during real-world operation. For example, visually dense interfaces may amplify cognitive interference, while increased cognitive load can in turn affect movement planning and execution efficiency. Such coupled effects are particularly evident in safety-critical domains \cite{bassano2020perception}, where operators must rapidly integrate perception, decision-making, and action under constrained conditions. Despite these insights, most existing studies examine these factors in isolation and lack a unified framework to quantify their combined impact on operator action time \cite{wu2015framework}. This limitation underscores the need for integrative modeling approaches that can capture the joint influence of interface characteristics on human performance, providing a foundation for ergonomic evaluation and interface design.

\subsection{Approaches to predicting human performance} \label{Approaches to predicting human performance}

A wide range of approaches have been developed to predict human performance in human–machine systems, particularly in terms of task completion time, workload, and error likelihood \cite{stowers2017framework}. One major category is empirical and experiment-based methods, which rely on controlled laboratory studies, simulator experiments, or field observations to measure operator behavior. These approaches provide high-fidelity data and have been widely used to evaluate interface usability and task efficiency \cite{pool2012evaluating}. However, they are often resource-intensive, requiring substantial time, specialized equipment, and access to trained participants. As a result, their applicability is limited in early-stage design, where rapid evaluation of multiple interface alternatives is needed \cite{dewan2018pediatric}.

Another line of research focuses on data-driven predictive models, including statistical regression and machine learning approaches \cite{ding2020comparison}. These methods can capture complex relationships between input features and performance outcomes and often achieve strong predictive accuracy when sufficient data are available \cite{hassija2024interpreting}. Nevertheless, such models typically function as black-box predictors, offering limited interpretability regarding the underlying cognitive and perceptual processes \cite{concone2024adverspam}. This lack of transparency makes it difficult to understand how specific interface characteristics contribute to performance changes, thereby reducing their usefulness for ergonomic analysis and design-oriented decision-making \cite{jayarajah2019occupational}.

Cognitive modeling approaches provide an alternative by explicitly representing the mechanisms underlying human behavior. Models based on frameworks such as ACT-R, GOMS, and Fitts’ law have been used to estimate task execution time by decomposing interaction into perceptual, cognitive, and motor components \cite{fum2007cognitive}. These approaches offer strong interpretability and theoretical grounding, making them well suited for analyzing human–machine interaction \cite{xiao2026implementation}. However, existing implementations often rely on manual task decomposition and predefined rules, and are rarely directly driven by quantifiable interface-level features. Consequently, there remains a gap in developing models that combine the interpretability of cognitive architectures with the ability to predict performance directly from measurable interface characteristics, particularly in complex and safety-critical digital environments \cite{kotseruba202040}.

\subsection{Research gap and positioning of this study} \label{Research gap and positioning of this study}

Despite substantial progress in understanding the effects of interface design on human performance and in developing predictive modeling approaches, several important gaps remain. First, existing studies have rarely established a direct and operational link between measurable interface characteristics and quantitative predictions of operator action time \cite{peruzzini2020exploring}. While prior work has identified factors such as visual salience, semantic complexity, and interaction distance as influential, these factors are typically examined in isolation and are not systematically integrated into a unified predictive framework \cite{wang2014isolation}. As a result, it remains difficult to translate interface-level design attributes into concrete estimates of task performance.

Second, there is a lack of modeling approaches that simultaneously achieve predictive capability and interpretability \cite{stiglic2020interpretability}. Empirical and data-driven methods can provide accurate estimates under specific conditions but often fail to reveal the underlying cognitive and perceptual mechanisms. In contrast, cognitive modeling approaches offer mechanistic insight but are commonly limited by their reliance on manual task decomposition and their weak connection to quantifiable interface features \cite{hsiao2024understanding}. This disconnect reduces their applicability in practical ergonomic evaluation and interface design, where both accuracy and interpretability are essential.

Third, existing studies have paid limited attention to user expertise differences when modeling interface-driven performance \cite{kaber2004effects}. In real-world systems, novice and experienced operators often exhibit distinct cognitive strategies, levels of automation, and sensitivity to interface complexity. Ignoring such differences may lead to biased or incomplete performance predictions, particularly in safety-critical environments where operator expertise plays a crucial role \cite{dreany2018safety}.

To address these gaps, the present study proposes a theory-guided and data-calibrated ACT-R-based framework that predicts operator action time directly from interface features. The proposed approach integrates measurable interface characteristics into a decomposed time model consisting of visual, motor, cognitive, and interaction components, enabling interpretable representation of human performance mechanisms \cite{giese2015neural}. In addition, a data-driven clustering method is employed to identify task types, and separate parameter sets are calibrated for novice and experienced users to capture expertise-related differences. By bridging interface-level design attributes with cognitively grounded performance modeling, this study positions itself as an integrative approach for ergonomic evaluation and interface design in complex human–machine systems \cite{kirkby2025artificial}.

\section{Methodology}

\subsection{Overview of methodology}\label{sec:overview}

The overall workflow is shown conceptually in four stages (Figure~\ref{framework}). First, interface- and trial-level data were extracted from the digital control room experiment. Each record contained a task step identifier, three recorded interface metrics (TS, SID, and IS), start and end coordinates of the action, user group, and response outcome. Second, successful trials were filtered and cleaned through a two-stage outlier-removal procedure, ensuring that extreme values caused by interruptions or logging artifacts were excluded. Third, task types were identified through K-Means clustering over the feature space defined by TS, SID, and cursor-travel distance, resulting in data-driven categorization of task difficulty. Fourth, ATIM parameters were calibrated separately for each user group and task type. As illustrated in Figure~\ref{framework}, the calibrated parameters are embedded into a theory-guided ACT-R-based time decomposition model, in which operator action time is represented as the sum of visual, cognitive, motor, and interaction components. The resulting model was then used for validation and for generating step-level ACT-R Lisp snippets, enabling integration with cognitive simulation.

\begin{center}
\includegraphics[width=1.0 \textwidth]{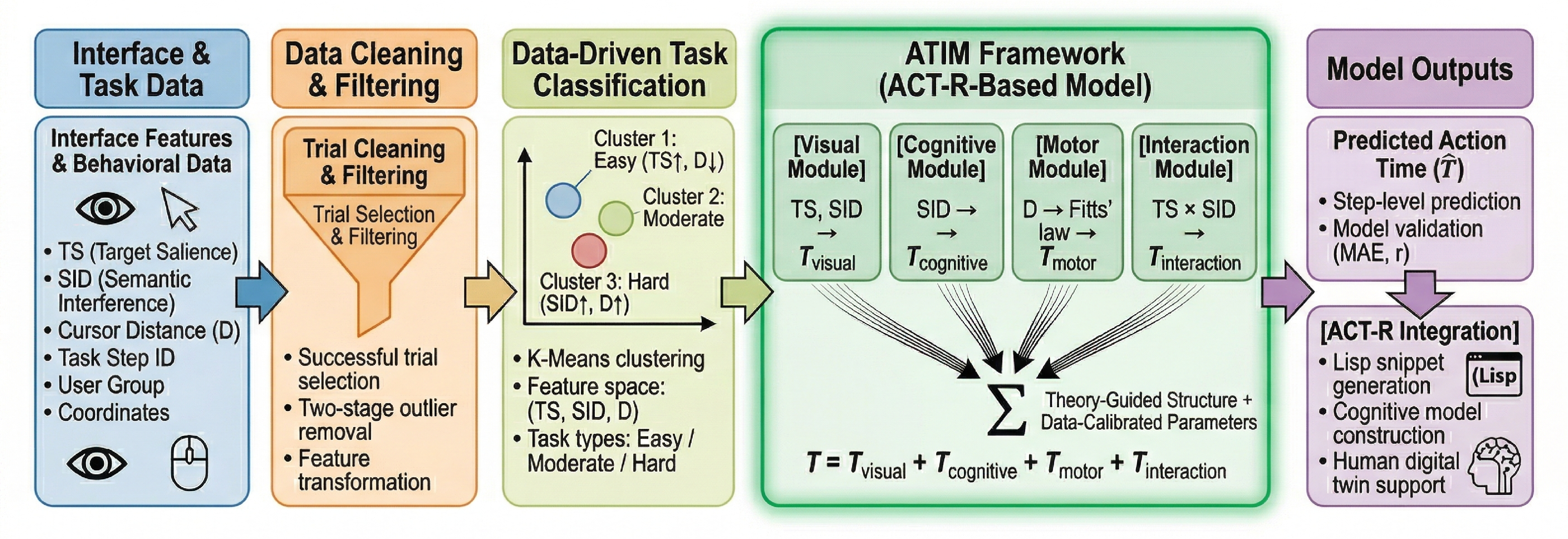}
\captionof{figure}{Overview of the ATIM framework for predicting operator action time from interface features. The framework integrates data processing, task classification, and a theory-guided ACT-R-based modeling structure to decompose action time into visual, cognitive, motor, and interaction components.}
\label{framework}
\end{center}

This design intentionally combines theory-driven structure and data-driven adaptation. As shown in Figure~\ref{framework}, the upper stages of the pipeline (data extraction, cleaning, and clustering) are data-driven, while the core ATIM model is constrained by established perceptual, cognitive, and motor principles. The timing equation itself is therefore theoretically grounded, whereas the coefficients and task categories are learned from data. This hybrid strategy preserves interpretability while avoiding the need to fix arbitrary thresholds or coefficients in advance.

\subsection{Experimental setup and task environment}

The experimental data were collected in a simulated digital nuclear control room environment, where participants performed structured verification tasks based on predefined operational procedures.

Each task required participants to read a textual procedure and verify corresponding parameters on a graphical control interface. The procedures consisted of multi-step instructions involving system identification, parameter lookup, and value comparison. These tasks reflect typical operator activities in digital control rooms, where information must be retrieved and validated across multiple interface regions.

\begin{center}
\includegraphics[width=0.8 \textwidth]{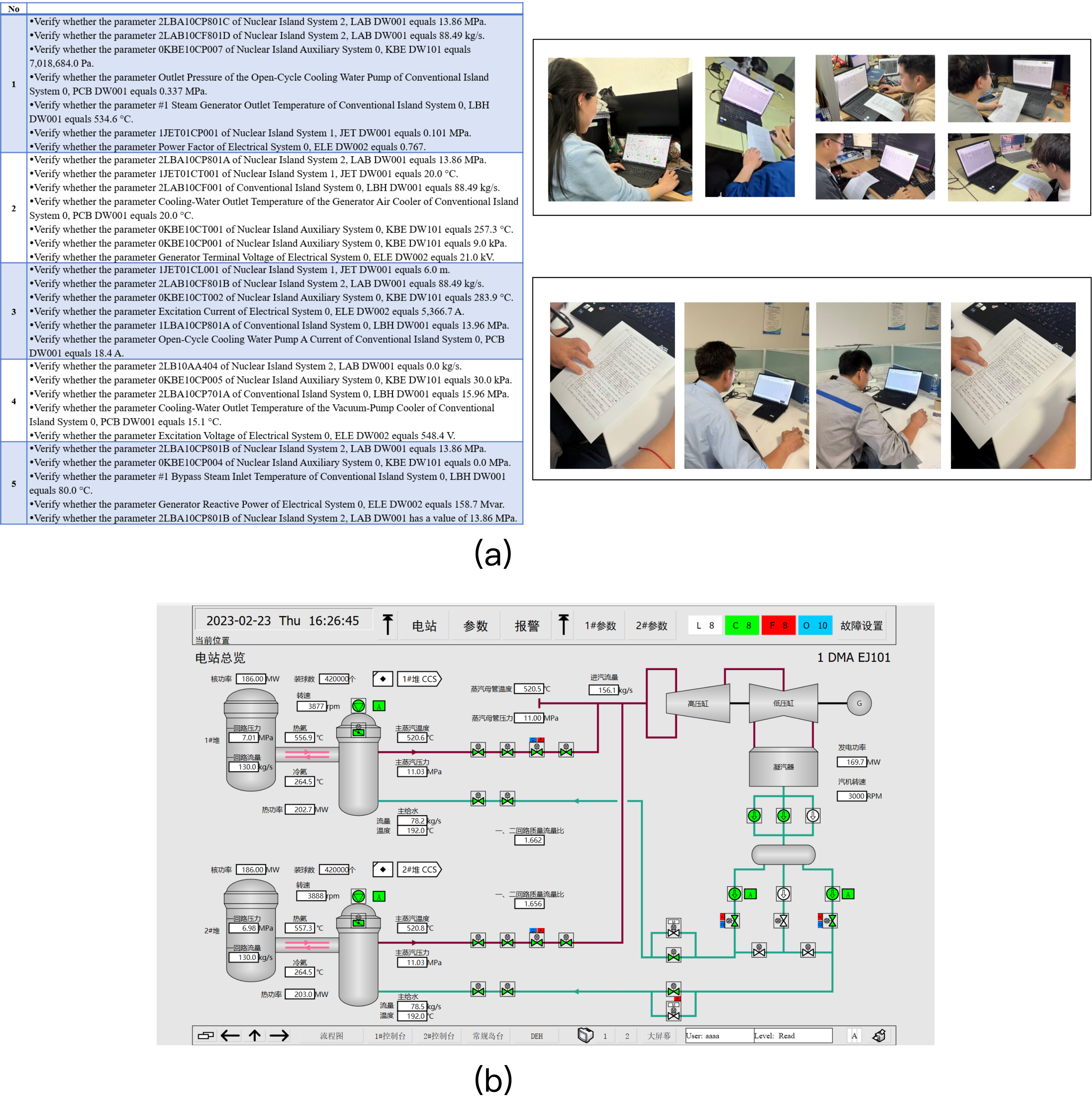}
\captionof{figure}{Experimental setup and task environment. (a) Participants performing procedure-based verification tasks using printed instructions. 
(b) Digital control room interface used for parameter verification. The experiment captures realistic human–machine interaction involving visual search, semantic processing, and motor execution.}
\label{exp}
\end{center}

The interface used in the experiment is shown in Fig.~\ref{exp} (a). It represents a digital control panel with multiple subsystems, numerical indicators, and control elements distributed across the screen. Participants interacted with the interface using a mouse, and their action trajectories and response times were recorded. An example of the task procedure and the experimental setting is shown in Fig.~\ref{exp} (b). Participants were required to follow the procedure sequentially while interacting with the interface. The setup captures the coupled perceptual–cognitive–motor demands of real-world operator tasks, including visual search, semantic interpretation, and cursor-based interaction.

Two participant groups were included: a student group representing novice users and an operator group representing experienced users. All participants completed the tasks under controlled conditions, and their interaction data were logged for subsequent modeling. This experimental design ensures that the collected data reflect realistic human–machine interaction processes in digital control rooms, providing a valid basis for interface-driven performance modeling.

\subsection{Interface feature representation}\label{sec:features}

Three interface features were used to characterize each task step. Target salience (TS) is defined as
\begin{equation}
\widetilde{TS} = \frac{N_{\text{target}}}{N_{\text{elements}}},
\end{equation}
where $N_{\text{target}}$ denotes the number of target elements (equal to 1 in the present tasks), and $N_{\text{elements}}$ represents the total number of visible elements on the interface. This formulation provides a simple proxy for the relative visual prominence of the target within the display.

From a human-factors perspective, TS reflects the difficulty of visual search. A lower TS value corresponds to a denser interface with more competing visual items, thereby increasing search complexity and prolonging the perceptual stage. Conversely, a higher TS indicates that the target occupies a larger proportion of the visual field and can be identified more rapidly. In the present dataset, TS was encoded as fractions (e.g., $1/4$, $1/13$), where smaller values indicate lower salience and thus greater search difficulty.

Semantic interference density (SID) is defined as
\begin{equation}
\widetilde{SID} = \frac{N_{\text{high-similarity}}}{N_{\text{total}}},
\end{equation}
where $N_{\text{high-similarity}}$ denotes the number of element pairs whose name similarity exceeds a predefined threshold, and $N_{\text{total}}$ represents the total number of possible element pairs on the interface.

SID captures the extent of semantic or functional similarity among interface elements, which contributes to cognitive interference during target identification. A higher SID indicates that multiple elements share similar labels or functional meanings, increasing the likelihood of confusion and requiring additional cognitive effort for discrimination. Unlike TS, which primarily affects the perceptual search stage, SID reflects difficulty at the cognitive stage, particularly in resolving ambiguity among competing alternatives.

Interaction span (IS) is defined as
\begin{equation}
\widetilde{IS} = \frac{\text{Euclidean Distance}(P_i, P_{i+1})}{\text{Distance}_{\text{longest}}},
\end{equation}
where $P_i$ and $P_{i+1}$ denote consecutive interaction points in the recorded trajectory, and $\text{Distance}_{\text{longest}}$ represents the maximum possible cursor travel distance across the interface.

IS provides a normalized measure of the spatial extent of motor execution between successive actions. Larger IS values indicate greater cursor travel and therefore increased motor effort. In the present dataset, IS was available as a recorded descriptor of movement demand. However, in the current implementation of ATIM, the motor component was modeled using Euclidean distance directly recovered from interface coordinates, which allows a more precise representation of movement cost in accordance with Fitts' law. IS is therefore retained as a conceptual descriptor of interaction span, while distance-based measures are used for quantitative modeling.

Taken together, TS, SID, and IS represent three complementary dimensions of interface-induced workload: perceptual load (TS), cognitive interference (SID), and motor demand (IS). This decomposition enables the proposed model to explicitly link interface characteristics to distinct stages of human information processing.

For calibration, the fraction-valued metrics were transformed into scalar values. Cursor travel distance $D$ was computed from the recorded begin and end coordinates:
\begin{equation}
D = \sqrt{(x_{\text{end}}-x_{\text{begin}})^2 + (y_{\text{end}}-y_{\text{begin}})^2}.
\end{equation}

This representation allows ATIM to map observable interface structure onto latent perceptual and motor demands in a transparent way.

\subsection{Rationale for the mathematical formulation}
The mathematical forms used in ATIM were designed to provide a first-order, mechanism-based approximation of the perceptual, cognitive, and motor processes involved in digital control room interaction. Rather than treating the equations as purely empirical regressors, each term was defined to correspond to a specific human-performance mechanism.

The visual component was formulated as 
$T_{\mathrm{visual}} = K_{\mathrm{visual}}(1/TS)(1+SID)$ 
to approximate the increase in visual search demand caused by reduced target salience and semantic distraction. Since $TS$ is defined as the ratio of target elements to all visible elements, its reciprocal $1/TS$ can be interpreted as an effective search-set size. A smaller $TS$ therefore corresponds to a larger number of competing elements and longer visual search time. The multiplier $(1+SID)$ was introduced to represent the amplifying effect of semantic interference on visual search. The constant term ensures that visual search time remains non-zero even when semantic interference is absent.

The cognitive component was defined as 
$T_{\mathrm{cognitive}} = K_{\mathrm{cognitive}} SID$, 
because semantic interference primarily affects the decision and confirmation stage rather than the motor execution stage. A higher $SID$ indicates that more interface elements share similar semantic or functional meanings, which increases the time required for discrimination, confirmation, and conflict resolution. A linear form was adopted as a first-order approximation to preserve interpretability and avoid over-parameterization given the limited experimental dataset.

The motor component followed a Fitts-like formulation:
$T_{\mathrm{motor}} = fitts_a + fitts_b \log_2(D/50+1)$.
Classical Fitts' law models \cite{bertucco2013fitts} movement time as a logarithmic function of the ratio between movement distance and target width. In the present dataset, cursor movement distance was available from recorded coordinates, whereas target width was not consistently recorded for all interface elements. Therefore, a fixed scaling constant of 50 pixels was used as an approximate reference width to normalize movement distance and construct a stable index of movement difficulty. This term was not intended to estimate pure motor execution time in a strict Fitts' law experiment, but to capture the distance-dependent component of cursor-based interaction in the digital control room interface.

Finally, the interaction component was introduced as 
$T_{\mathrm{interaction}} = \alpha SID(1/TS)$ 
to capture the coupled effect of low visual salience and high semantic interference. These two factors may not affect performance independently. When a target is visually inconspicuous and semantically similar distractors are present, operators may need to perform repeated search and confirmation, leading to disproportionately longer action time. The multiplicative term therefore represents an amplification mechanism beyond the additive visual and cognitive components.

\subsection{ACT-R-based time decomposition model}\label{sec:decomposition}

ATIM assumes that operator action time can be decomposed into four additive components corresponding to distinct stages of human information processing within the ACT-R framework:
\begin{equation}
T = T_{\text{visual}} + T_{\text{motor}} + T_{\text{cognitive}} + T_{\text{interaction}}.
\end{equation}

This decomposition reflects a structured mapping between interface-driven demands and underlying perceptual, cognitive, and motor processes. Each component captures a specific mechanism through which interface characteristics influence performance.

The additive structure is a first-order approximation that assumes sequential processing stages in ACT-R. While real interactions may involve overlap or feedback loops, this formulation provides a tractable and interpretable baseline.

\paragraph{Visual component.}
The visual search component is defined as
\begin{equation}
T_{\text{visual}} = K_{\text{visual}}\left(\frac{1}{TS}\right)(1 + SID),
\end{equation}
where $TS$ represents target salience and $SID$ denotes semantic interference density. This formulation is grounded in visual search theory, where low-salience targets require longer search times, and the presence of semantically similar distractors increases attentional competition. The multiplicative structure reflects the assumption that salience and interference jointly affect search efficiency rather than acting independently.

\paragraph{Motor component.}
The motor component follows a Fitts-like formulation:
\begin{equation}
T_{\text{motor}} = fitts_a + fitts_b \log_2\left(\frac{D}{50}+1\right),
\end{equation}
where $D$ is the cursor travel distance in pixels. This formulation captures the well-established relationship between movement time and distance in human motor control. The normalization constant (50 pixels) is introduced to align the scale of movement with the interface resolution used in the experiment, ensuring numerical stability and interpretability of the fitted parameters.

\paragraph{Cognitive component.}
The cognitive component is modeled as
\begin{equation}
T_{\text{cognitive}} = K_{\text{cognitive}} \cdot SID,
\end{equation}
which represents the additional time required for decision-making and conflict resolution. Higher semantic interference increases ambiguity among competing options, thereby prolonging cognitive processing time. This term captures decision latency that is not directly attributable to visual search or motor execution.

\paragraph{Interaction component.}
Finally, an interaction term is introduced to capture nonlinear coupling between low salience and high interference:
\begin{equation}
T_{\text{interaction}} = \alpha \cdot SID \left(\frac{1}{TS}\right).
\end{equation}
This term accounts for situations in which the combined effect of poor visual distinguishability and semantic similarity leads to disproportionately large delays. Such coupling effects are frequently observed in complex interfaces, where multiple sources of difficulty amplify each other.

\paragraph{Full prediction model.}
The complete prediction function is therefore given by
\begin{equation}
\hat{T} = K_{\text{visual}}\left(\frac{1}{TS}\right)(1 + SID) + fitts_a + fitts_b \log_2\left(\frac{D}{50}+1\right) + K_{\text{cognitive}} \cdot SID + \alpha \cdot SID\left(\frac{1}{TS}\right).
\end{equation}

This formulation provides a direct mapping from measurable interface features to predicted action time, while preserving interpretability at the level of individual cognitive and perceptual processes.

\paragraph{Parameter constraints.}
The model parameters were constrained to remain within psychologically plausible ranges: $K_{\text{visual}} \ge 0.05$, $K_{\text{cognitive}} \ge 0.02$, $fitts_a \ge 0.10$, $fitts_b \ge 0.05$, and $\alpha \ge 0$. These constraints ensure that each component contributes meaningfully to the overall prediction and prevent degenerate solutions in which theoretically necessary processes collapse to zero.

From a modeling perspective, these bounds also stabilize the optimization process and maintain consistency with empirical findings in human performance research, where perceptual, cognitive, and motor processes exhibit non-zero baseline costs.

\subsection{Data processing and task classification}\label{sec:classification}

The behavioral dataset contained 567 successful trials (i.e., trials with successful completion code $S0=1$) collected from a simulated digital nuclear control room task. The participants were divided into two recorded groups in the data table: a student group and an operator group.

Because response-time data in simulator environments may include lapses, interruptions, or logging artifacts, a two-stage outlier treatment was applied. In Stage 1, response times were standardized within each $(step, group)$ condition and trials with $|Z| > 2.5$ were excluded. In Stage 2, a global standardization was applied to the retained data and trials with extreme global deviations were removed. This process excluded 19 trials (3.4\%) and retained 548 trials for modeling.

Task types were not assigned by hand-crafted thresholds. Instead, K-Means clustering was performed on the standardized feature vector $(TS, SID, D)$. The script compared candidate solutions over $K=2$ to $6$ using inertia drops, then enforced a minimum of three clusters and produced a final solution with $K=3$. Cluster centroids were ranked by a difficulty score proportional to $(1/TS)(1+SID)+D/500$, after which the clusters were labeled as easy, moderate, and hard. Table~\ref{tab:clusters} reports the centroid values and the number of retained trials in each class.

\begin{table}[htbp]
\caption{Data-driven task classes identified by K-Means clustering after outlier removal.}
\label{tab:clusters}
\centering
\begin{tabular}{lrrrr}
\toprule
Task type & Mean TS & Mean SID & Mean distance (px) & Trials \\
\midrule
Easy & 0.249 & 0.003 & 489.3 & 187 \\
Moderate & 0.217 & 0.944 & 636.0 & 143 \\
Hard & 0.078 & 0.218 & 794.8 & 218 \\
\bottomrule
\end{tabular}
\end{table}

\subsection{Parameter calibration}\label{sec:calibration}
Parameters were calibrated separately for each combination of user group and task type. Let $\theta = \{K_{\text{visual}}, K_{\text{cognitive}}, fitts_a, fitts_b, \alpha\}$ denote the parameter vector for a given subgroup. The optimization objective minimized mean squared error between predicted and observed trial time:
\begin{equation}
\theta^{*} = \arg\min_{\theta} \frac{1}{n}\sum_{i=1}^{n} \left(\hat{T}_i(\theta) - T_i\right)^2.
\end{equation}

Bounded optimization was implemented with the L-BFGS-B algorithm. Calibration was conducted separately for student and operator data. After calibration, the model was validated by averaging predictions and observations at the step level within each group and comparing them using mean absolute error (MAE), root mean square error (RMSE), and Pearson correlation.

In addition to prediction, the calibrated parameters were embedded into an ACT-R code generation routine that produced Lisp snippets for specific interface steps. This connection is practically important because it allows the timing model to support future cognitive model assembly rather than serving only as an offline regression tool.

\subsection{Generalization evaluation}\label{sec:validation}

To evaluate the generalization capability of the proposed framework, a stratified 70/30 split was applied to the cleaned dataset. The split was performed within each subgroup defined by user group (student vs.\ operator) and task type (easy, moderate, hard), ensuring balanced representation across training and validation sets.

Model parameters were calibrated exclusively on the training subset (70\%), and the learned parameter sets were then applied to the held-out validation subset (30\%) without further adjustment. This procedure ensured that both parameter estimation and performance evaluation were conducted without information leakage.

Performance was evaluated at both trial level and step level using mean absolute error (MAE), root mean square error (RMSE), and Pearson correlation coefficient ($r$).

\section{Results}\label{sec:results}

\subsection{Overall prediction performance}

To evaluate the generalization capability of ATIM, a stratified 70/30 split was applied to the cleaned dataset. Model parameters were calibrated on the training subset and subsequently evaluated on the held-out validation subset without further adjustment. Figure~\ref{fig:pred_vs_obs} illustrates the relationship between predicted and observed step-level action times on the validation set. The data points are closely distributed along the diagonal line, indicating strong agreement between predictions and measurements. This result demonstrates that the model effectively preserves both the absolute magnitude of action time and the relative ordering across task steps.

\begin{center}
\includegraphics[width=0.5 \textwidth]{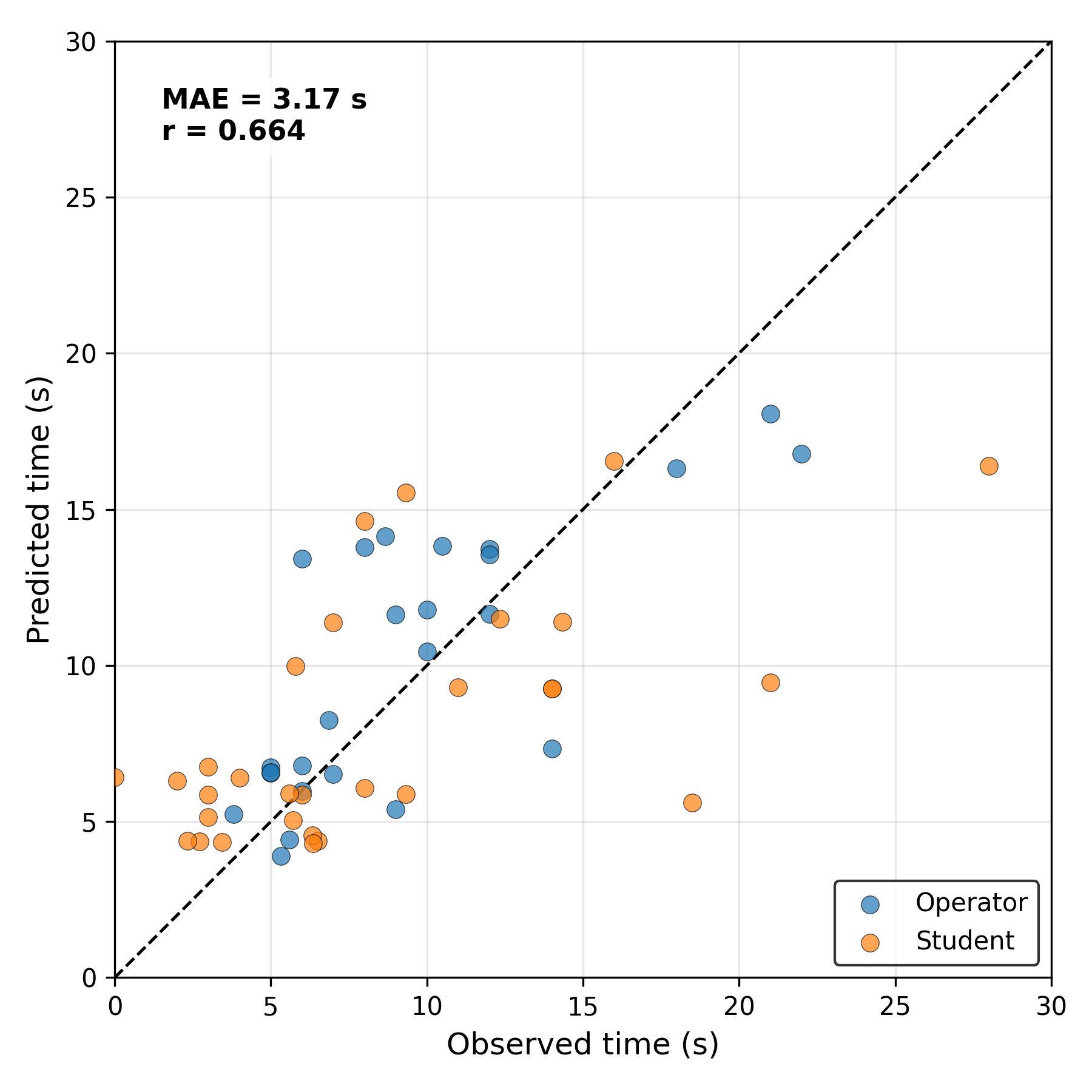}
\captionof{figure}{Predicted versus observed action time at the step level on the validation set. The dashed line indicates perfect agreement ($y=x$).}
\label{fig:pred_vs_obs}
\end{center}

To further examine potential systematic bias, the distribution of prediction residuals is shown in Figure~\ref{fig:residuals}. The residuals are approximately centered around zero and exhibit no evident skewness or long-tail behavior. This indicates that the model does not systematically overestimate or underestimate action time and provides stable predictions across different task conditions.

\begin{center}
\includegraphics[width=0.8 \textwidth]{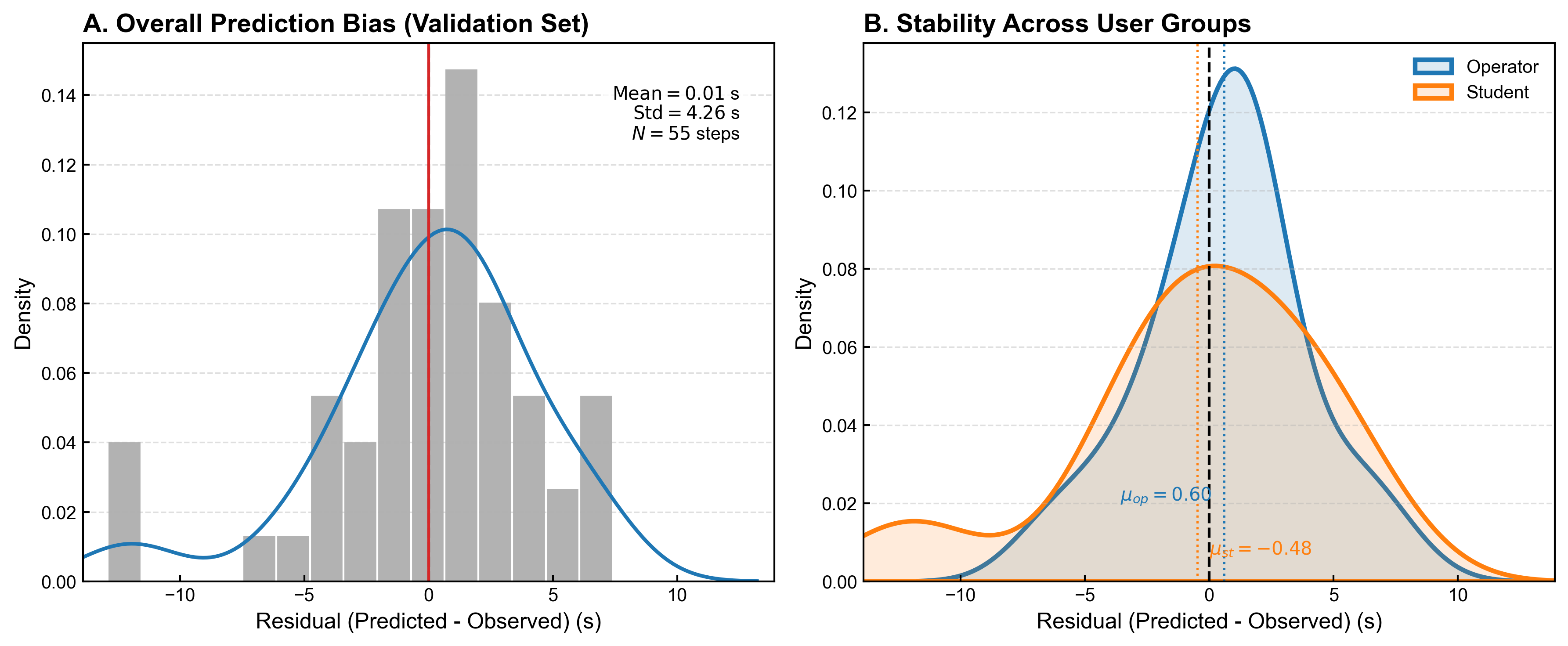}
\captionof{figure}{Distribution of prediction residuals on the validation set. Residuals are centered around zero, indicating no systematic bias.}
\label{fig:residuals}
\end{center}

Table~\ref{tab:generalization} summarizes the performance across the full dataset, training subset, and validation subset. At the step level, ATIM achieved a mean absolute error (MAE) of 3.17~s and a root mean square error (RMSE) of 4.26~s on the validation set, with a Pearson correlation coefficient of $r=0.664$. Compared with the full-data evaluation (MAE~=~2.72~s, $r=0.729$), the observed performance degradation is moderate, suggesting that the model retains substantial predictive capability under out-of-sample conditions.

At the trial level, the validation MAE (3.16~s) is comparable to that of the full dataset (3.21~s), indicating that the calibrated parameters capture stable relationships between interface features and action time. Although the correlation coefficient decreases from 0.637 to 0.567, this reduction is expected due to the smaller sample size and increased variability in the held-out subset.

\begin{table}[htbp]
\caption{Prediction performance under stratified 70/30 split.}
\label{tab:generalization}
\centering
\begin{tabular}{lccc}
\toprule
Metric & Full data & Training & Validation \\
\midrule
Trial MAE (s) & 3.21 & 3.27 & 3.16 \\
Trial RMSE (s) & 4.41 & 4.53 & 4.23 \\
Trial $r$ & 0.637 & 0.653 & 0.567 \\
Step MAE (s) & 2.72 & 3.14 & 3.17 \\
Step RMSE (s) & 3.62 & 4.01 & 4.26 \\
Step $r$ & 0.729 & 0.700 & 0.664 \\
\bottomrule
\end{tabular}
\end{table}

Overall, these results demonstrate that ATIM maintains robust predictive performance under out-of-sample evaluation. The moderate increase in error and slight reduction in correlation are consistent with expected generalization behavior, indicating that the model captures stable relationships between interface characteristics and operator action time rather than relying on overfitting. Importantly, this level of performance is achieved while preserving interpretability, as each component of the model remains grounded in a distinct perceptual, cognitive, or motor mechanism.

\subsection{Group-level results}

The model exhibits consistently better performance for the operator group than for the student group. Specifically, the operator model achieves an MAE of 2.30~s and RMSE of 3.16~s, compared to 3.09~s and 3.98~s for the student model. In addition, the operator group shows a stronger correlation with observed step time ($r=0.764$ vs.\ $r=0.711$), indicating more reliable preservation of relative task difficulty.

\begin{table}[htbp]
\caption{Validation performance by user group.}
\label{tab:groupresults}
\centering
\begin{tabular}{lrrrr}
\toprule
Group & Steps & MAE (s) & RMSE (s) & $r$ \\
\midrule
Student & 37 & 3.09 & 3.98 & 0.711 \\
Operator & 32 & 2.30 & 3.16 & 0.764 \\
Overall & 69 & 2.72 & 3.62 & 0.729 \\
\bottomrule
\end{tabular}
\end{table}

These quantitative differences are further supported by the distribution of prediction errors, where the operator group exhibits both lower error levels and reduced variability compared to the student group (Fig.~\ref{fig:group}). This indicates that the model provides more stable and accurate predictions for experienced operators.

\begin{center}
\includegraphics[width=0.8 \textwidth]{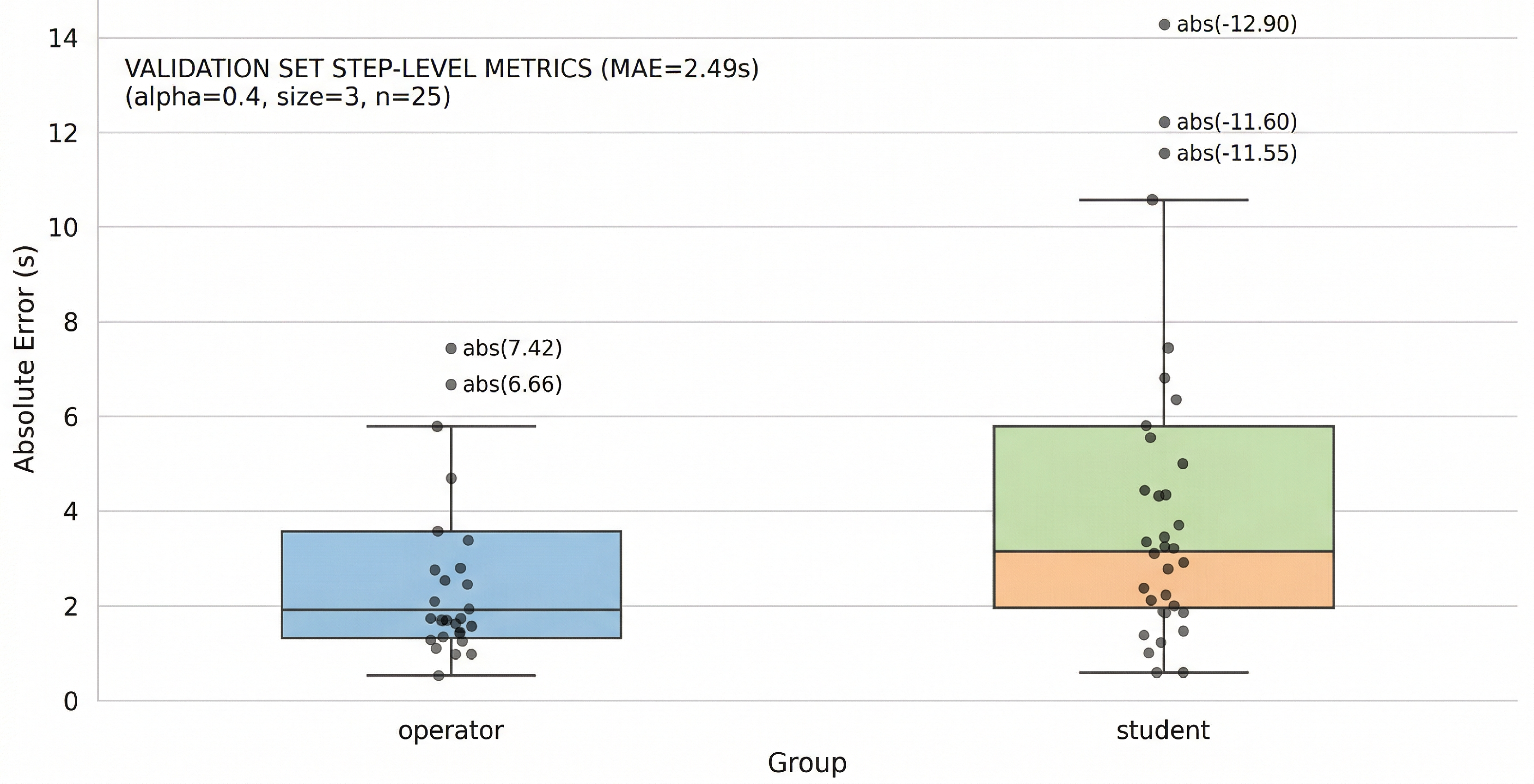}
\captionof{figure}{Comparison of prediction error (MAE) between student and operator groups on the validation set.}
\label{fig:group}
\end{center}

The observed difference can be explained by the underlying behavioral characteristics of the two groups. Operator behavior tends to be more structured and consistent, which aligns well with the theory-guided decomposition of perceptual, cognitive, and motor processes in ATIM. In contrast, the student group exhibits greater variability in action time, likely due to less stable visual search strategies, increased decision uncertainty, and occasional hesitation during interaction. These factors are not fully captured by the current feature representation and therefore lead to larger prediction errors.

Further insight can be obtained from the calibrated parameter values summarized in Table~\ref{tab:params}. Several patterns are noteworthy. First, the student easy-task model exhibits substantially larger $K_{\text{visual}}$, $K_{\text{cognitive}}$, and $\alpha$ values compared to the operator model, indicating higher perceptual and cognitive costs even in relatively simple tasks. This is consistent with less efficient visual search and decision-making processes in the student group.

Second, moderate tasks in both groups are characterized by larger $fitts_a$ values, suggesting that action initiation plays a more prominent role under high interference conditions. This may reflect cautious or stepwise interaction strategies when users are faced with semantically ambiguous interface elements.

Third, the operator model in hard tasks shows the highest $fitts_b$ value, indicating increased sensitivity to movement distance. This suggests that, under demanding conditions, motor execution becomes a dominant factor for experienced operators, who otherwise maintain efficient perceptual and cognitive processing.

\begin{table*}[htbp]
\caption{Calibrated parameter values by group and task type.}
\label{tab:params}
\centering
\begin{tabular}{l l r r r r r r}
\toprule
Group & Task type & $K_{\text{visual}}$ & $K_{\text{cognitive}}$ & $fitts_a$ & $fitts_b$ & $\alpha$ & $n$ \\
\midrule
Student & Easy & 1.036 & 5.000 & 0.100 & 0.050 & 5.000 & 152 \\
Student & Moderate & 0.064 & 0.020 & 1.500 & 0.899 & 0.000 & 88 \\
Student & Hard & 0.266 & 0.020 & 0.100 & 1.109 & 0.314 & 141 \\
Operator & Easy & 0.460 & 0.745 & 0.100 & 1.022 & 2.280 & 35 \\
Operator & Moderate & 0.077 & 0.020 & 1.500 & 0.924 & 0.000 & 55 \\
Operator & Hard & 0.050 & 0.020 & 0.100 & 2.553 & 0.227 & 77 \\
\bottomrule
\end{tabular}
\end{table*}

Overall, these results indicate that ATIM not only captures overall performance trends but also reflects meaningful differences in cognitive and interaction strategies between user groups. The framework therefore provides a mechanism-based interpretation of how expertise influences the relative contribution of perceptual, cognitive, and motor processes to task execution time.

\subsection{Error analysis}

Although the overall predictive performance of ATIM is encouraging, non-negligible errors remain for a subset of steps. Among the 69 step-level observations, 14 steps exhibit an absolute prediction error greater than 5~s, indicating that certain task conditions are not fully captured by the current model formulation.

A closer inspection of these cases reveals several representative patterns. The largest underestimation on the student side occurs at Step~211, where the observed mean time is 25.33~s compared to a predicted value of 15.66~s. A similar underestimation is observed for the operator group at Step~212 (25.50~s observed vs.\ 17.50~s predicted). In contrast, a notable overestimation is observed at student Step~124, where the model predicts 11.06~s for an observed time of only 2.33~s.

To better understand the distribution of errors across steps, Fig.~\ref{fig:error_step} presents the absolute prediction error for each step. While most steps exhibit relatively small deviations, a limited number of outliers contribute disproportionately to the overall error. This suggests that model inaccuracies are concentrated in specific task conditions rather than being uniformly distributed.

\begin{center}
\includegraphics[width=0.8 \textwidth]{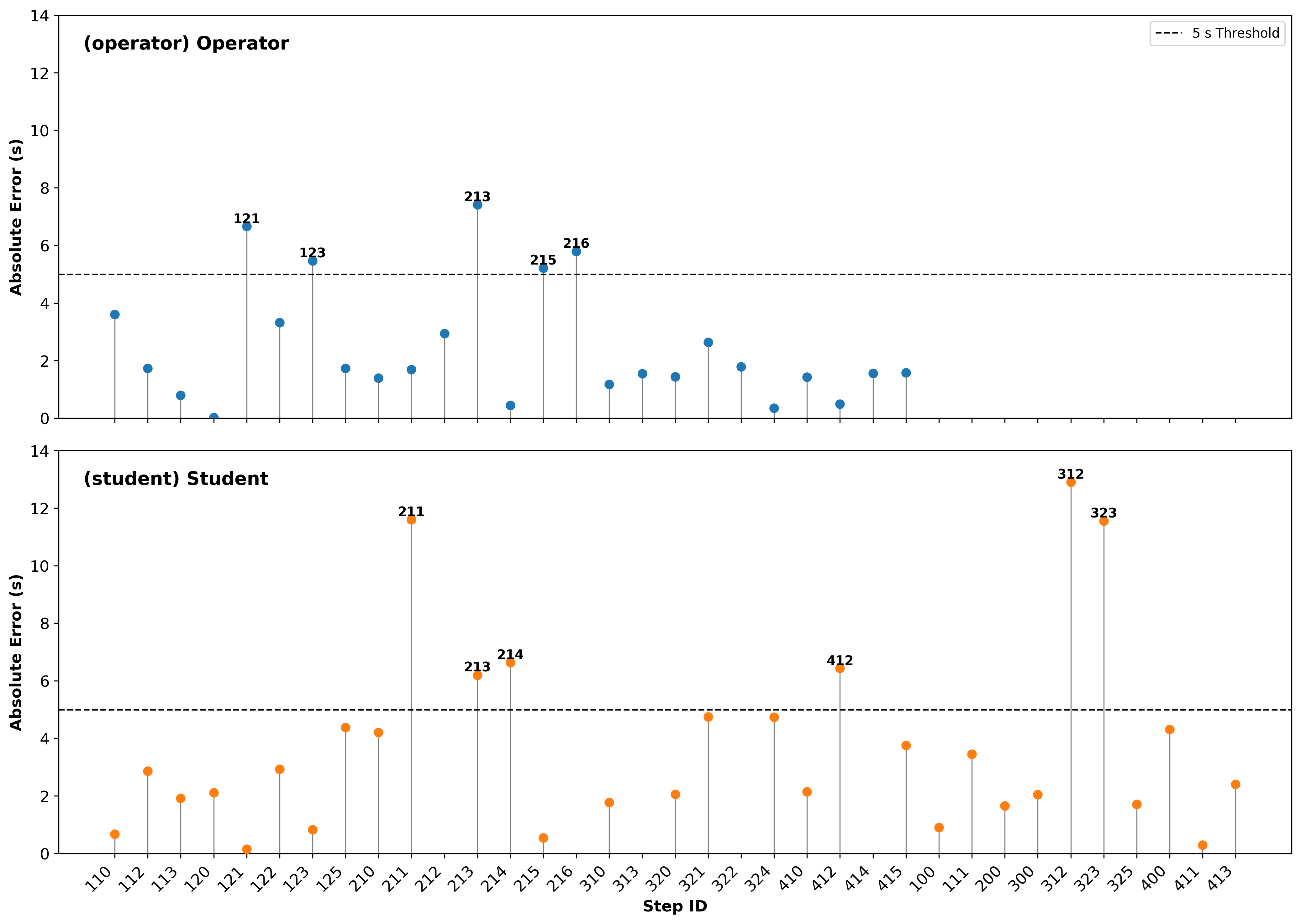}
\captionof{figure}{Absolute prediction error for each step on the validation set. A small number of steps exhibit significantly larger errors.}
\label{fig:error_step}
\end{center}

These deviations suggest that human action time in complex interfaces may not always follow a single-pass perception–cognition–action sequence, but can involve iterative loops and strategy shifts.

These high-error cases suggest that certain processes are not fully captured by the current time decomposition. In particular, steps involving hesitation, verification, repeated visual search, or multiple sub-actions within a single logged step may introduce additional delays that are not explicitly modeled. Such effects violate the assumption of a single-pass perception–cognition–action sequence and instead reflect more complex interaction dynamics.

Furthermore, several high-error steps are associated with larger within-group variability, indicating that prediction error may also be influenced by between-subject differences. This suggests that individual differences in strategy, experience, or attention allocation could contribute to residual variance beyond what is captured by interface features alone.

Overall, the error analysis indicates that while ATIM provides a robust first-order approximation of operator action time, its accuracy is limited in scenarios involving complex or multi-stage interaction patterns. Future work may incorporate additional behavioral indicators or hierarchical task representations to better capture such effects.

\section{Discussion}\label{sec:discussion}

\subsection{Human factors interpretation}

The results support the core proposition of ATIM: interface characteristics can be translated into interpretable predictions of operator action time. From a human-factors perspective, this is particularly important because it establishes an explicit mapping between measurable interface attributes and underlying perceptual, cognitive, and motor processes, rather than treating execution time as a purely empirical outcome.

The cluster structure derived from the data is especially revealing. Easy tasks combine relatively high target salience, minimal semantic interference, and short movement paths, indicating that performance is primarily supported by efficient visual localization and low decision ambiguity. In contrast, hard tasks are dominated by very low salience and longer movement distances, suggesting that increased execution time arises from the combined burden of difficult visual search and elevated motor demand. Moderate tasks form a distinct interference-driven class rather than lying midway between easy and hard. This indicates that semantic competition constitutes an independent source of difficulty in digital control room tasks.

\begin{center}
\includegraphics[width=0.6 \textwidth]{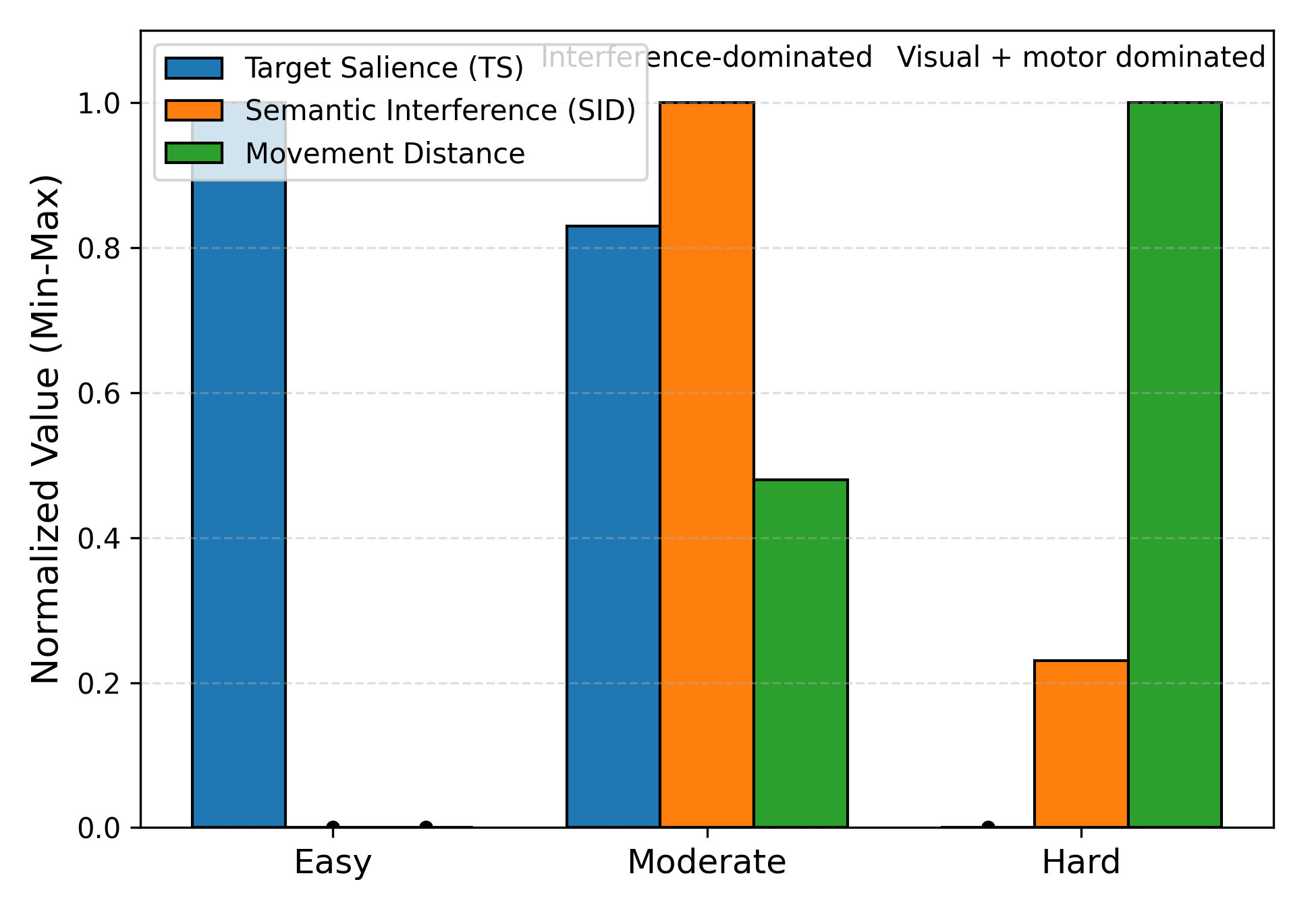}
\captionof{figure}{Normalized feature profiles of the three task classes identified by K-Means clustering. Moderate tasks are characterized by high semantic interference, whereas hard tasks are dominated by low target salience and long movement distance.}
\label{fig:task_profile}
\end{center}

The distinct feature profiles across task classes, shown in Fig.~\ref{fig:task_profile}, further support this interpretation. Easy tasks are characterized by high salience and negligible interference, whereas moderate tasks are dominated by semantic competition. In contrast, hard tasks exhibit low salience combined with large movement distance, indicating a shift from cognitive-dominant difficulty to perceptual–motor constraints. Importantly, this pattern demonstrates that task difficulty in digital control rooms is not a single continuum but emerges from different combinations of interface-related factors.

The parameter estimates provide additional insight into how these interface demands unfold across cognitive stages. Large visual and interaction coefficients in easy tasks indicate that even seemingly straightforward displays remain sensitive to subtle variations in salience and local distraction. This suggests that “easy” tasks are not necessarily robust and can still incur measurable performance degradation under minor interface perturbations.

In contrast, the high distance sensitivity observed in hard operator tasks indicates that spatial layout becomes a dominant bottleneck under demanding conditions. This finding is consistent with the notion that, as perceptual and cognitive processing become more efficient with expertise, motor execution increasingly limits performance. Consequently, long cursor travel and dispersed control layouts can significantly delay task completion even for experienced operators.

From a design perspective, these results imply that different ergonomic strategies are required for different task classes. Reducing semantic ambiguity is critical for interference-dominated tasks, improving visual salience is essential for perceptually challenging tasks, and minimizing movement distance is key for motor-intensive operations. Taken together, these findings highlight that operator performance is shaped by the interaction of multiple interface-driven mechanisms rather than by overall workload alone.

\subsection{Student vs.\ operator differences}

The distinction between student and operator models was not only statistically useful but also informative at the cognitive-mechanism level. Students exhibited higher prediction error and larger parameter values in easy tasks, particularly for the cognitive and interaction terms. This pattern suggests that novice users incur additional decision and confirmation costs even under relatively simple interface conditions. In terms of cognitive processing, this may reflect a greater reliance on explicit reasoning and verification, rather than efficient cue-based recognition.

By contrast, the operator group was more accurately predicted and showed lower cognitive weighting in easy conditions. This indicates that experienced operators may rely more on learned associations or perceptual shortcuts, allowing them to map interface cues to actions with reduced deliberation. Such behavior is consistent with the formation of stable cognitive representations and chunking mechanisms, which reduce the need for step-by-step decision processing.

Interestingly, the two groups showed similar parameter patterns in the moderate task class. In both cases, $fitts_a$ approached its upper bound, while $K_{\text{cognitive}}$ was reduced to its lower bound. This suggests that under strong semantic interference, both novice and experienced users may adopt more cautious or sequential interaction strategies. In such situations, execution time is less dominated by explicit cognitive weighting and more influenced by the need to resolve ambiguity through iterative interaction.

This convergence also highlights a limitation of the current feature representation. When interference becomes the dominant source of difficulty, user behavior may involve repeated visual search, re-checking, or corrective actions that are not fully captured by the present decomposition. As a result, the model shifts explanatory weight toward motor initiation terms, even though the underlying difficulty remains cognitive in nature.

Overall, these findings indicate that expertise primarily affects how interface information is processed rather than which interface features are relevant. Novice users tend to exhibit higher cognitive load and interaction sensitivity even in simple conditions, whereas experienced operators shift toward more efficient perceptual processing but may become constrained by motor factors in complex tasks. This highlights that operator performance differences arise from changes in the dominant processing stage, rather than from a uniform reduction in task difficulty.

\subsection{Implications for interface design}

ATIM provides practical value for interface design in two complementary ways. First, it enables rapid comparison of alternative interface configurations without requiring time-consuming human-subject experiments. By directly linking measurable interface features to predicted action time, designers can estimate the performance impact of changes such as reducing target salience, increasing the number of semantically similar controls, or dispersing sequential actions across distant regions of the screen. This makes ATIM suitable as a preliminary screening tool during early-stage interface design.

Second, the decomposed structure of the model allows delays to be attributed to specific perceptual, cognitive, or motor processes. Unlike black-box predictive models, which provide limited diagnostic insight, ATIM can indicate whether performance degradation arises primarily from visual search difficulty, semantic ambiguity, or movement-related constraints. This interpretability makes the model more actionable for iterative design refinement, as targeted modifications can be applied to the most influential component.

The results of the present study further suggest that different classes of interface tasks require distinct ergonomic strategies. For perceptually driven tasks, improving target salience through enhanced contrast, color coding, or grouping can significantly reduce visual search time. For interference-dominated tasks, minimizing semantic similarity among neighboring controls and providing clearer labeling or contextual cues is critical to reducing decision ambiguity. For motor-intensive tasks, especially those involving experienced operators, reducing cursor travel by spatially co-locating sequentially coupled controls can substantially improve execution efficiency.

Importantly, these strategies should not be treated as interchangeable. The findings indicate that task difficulty in digital control rooms arises from different combinations of interface features rather than from a single underlying factor. As a result, effective interface optimization requires identifying the dominant source of delay for a given task and applying the corresponding design intervention. This reinforces the central contribution of ATIM as a mechanism-based tool for supporting evidence-informed interface design in safety-critical systems.

\subsection{Methodological contribution}

Methodologically, this study contributes a hybrid approach that bridges cognitive theory and automated modeling. Instead of manually specifying timing assumptions for each interface configuration, the proposed framework infers the temporal consequences of interface features through data-calibrated parameters, while preserving a theory-grounded decomposition structure. This combination allows the model to remain interpretable in terms of perceptual, cognitive, and motor processes, while avoiding the rigidity and scalability limitations of fully hand-crafted cognitive models.

Compared with traditional ACT-R modeling workflows, which often require extensive manual encoding of task-specific rules and timing parameters, ATIM shifts part of the modeling effort from rule specification to feature-based estimation. In this sense, the framework can be viewed as a semi-automated layer on top of cognitive architecture modeling, where interface descriptors serve as inputs and calibrated parameters serve as reusable components. This significantly reduces the effort required to adapt the model to new interface variants or task scenarios, while maintaining a transparent link between model behavior and human-performance theory.

This methodological contribution is particularly relevant to the broader goal of developing human digital twins for safety-critical systems. In the current implementation, ATIM includes a routine that automatically generates step-level ACT-R Lisp fragments from interface metrics and task descriptions. The calibrated timing parameters are embedded into these snippets, enabling rapid construction of executable cognitive models without fully manual specification. As a result, ATIM functions not only as a predictive model, but also as a reusable timing layer that can support large-scale cognitive simulation.

More broadly, the proposed framework illustrates a pathway toward integrating cognitive architectures with data-driven calibration and engineering-oriented automation. This integration addresses a long-standing gap between theoretically grounded models, which are often difficult to scale, and purely data-driven approaches, which lack interpretability. By combining these paradigms, ATIM provides a foundation for scalable, interpretable, and application-oriented cognitive modeling in complex human–machine systems.

Compared with traditional ACT-R modeling, ATIM reduces manual modeling effort while preserving interpretability, and compared with machine learning models, it provides explicit mechanism-level explanations.

\section{Limitations and Future Work}

Several limitations should be noted. First, although a stratified train–validation split was used to assess generalization, the dataset is still limited to a single experimental environment. Further evaluation on independent datasets and different interface configurations is needed to fully assess external validity. Second, the current data originate from a specific digital nuclear control room task environment, which limits generalizability across plants, interface styles, and operational contexts.

Third, the feature set remains compact. While TS, SID, and movement distance capture important perceptual and motor demands, they do not explicitly encode factors such as temporal pressure, memory load, step-to-step dependency, or uncertainty during decision confirmation. This omission likely contributes to the remaining large errors on a subset of steps. Fourth, the current formulation treats each logged step as a single action episode, whereas some steps may contain retries, verification loops, or latent subgoals.

Future work should therefore expand the feature representation, incorporate sequential context and learning effects, and evaluate the model on independent datasets. Another promising direction is to link ATIM more tightly with large-language-model-based ACT-R code generation, so that task descriptions, interface features, and timing parameters can be translated into larger assembled cognitive models with less manual intervention.

\section{Conclusions}\label{sec:conclusion}

This study presented ATIM, an ACT-R-based task interface model for predicting operator action time in digital nuclear control rooms. By decomposing action time into visual, motor, cognitive, and interaction components and calibrating the associated parameters using behavioral data, ATIM establishes an interpretable link between interface characteristics and operator performance.

The empirical results demonstrate that the proposed framework achieves reliable predictive accuracy while maintaining interpretability. Under a stratified 70/30 validation scheme, ATIM achieved a step-level MAE of 3.17~s and a correlation coefficient of 0.664 on unseen data, indicating its capability to generalize beyond the calibration dataset. In addition to overall performance, the model reveals meaningful differences between student and operator groups, suggesting that expertise influences the relative contribution of perceptual, cognitive, and motor processes rather than uniformly reducing task difficulty.

More importantly, ATIM contributes a mechanism-based perspective on interface-driven human performance. The results show that task difficulty in digital control rooms arises from distinct combinations of interface features, including target salience, semantic interference, and movement distance. By explicitly modeling these factors, the framework provides actionable insights for interface design, enabling targeted improvements based on the dominant source of delay.

Beyond prediction, ATIM also provides a foundation for integrating cognitive modeling into practical engineering workflows. The step-level decomposition supports automatic generation of ACT-R Lisp snippets and offers a scalable pathway toward human digital twin modeling in safety-critical systems. This integration of interpretable modeling and data-driven calibration represents a step toward bridging cognitive theory and real-world system design.

Several limitations should be acknowledged. The current model relies on a limited set of interface features and assumes a relatively simple decomposition structure, which may not fully capture complex behaviors such as repeated search, hesitation, or strategy switching under high uncertainty. Future work will extend the feature representation, incorporate richer behavioral signals, and explore adaptive or hierarchical modeling approaches to better represent dynamic human–machine interaction.

In summary, ATIM demonstrates that theory-guided, interface-driven modeling can provide both accurate prediction and meaningful interpretation of operator behavior. This highlights its potential as a practical tool for ergonomic evaluation, interface optimization, and the development of cognitively grounded digital twins in high-risk industrial systems.

\printcredits

\bibliographystyle{cas-model2-names}

\bibliography{cas-refs}

\end{document}